\documentclass[prd,aps,showpacs,twocolumn]{revtex4}
\usepackage{graphics}
\begin{document}
\title{Hadronic decays from the lattice}

\author{Chris Michael}

\affiliation{Theoretical Physics Division, Dept. Math. Sci., University
of Liverpool,  Liverpool L69 7ZL, UK.}

\begin{abstract}
 I review the lattice QCD approach to determining hadronic decay
transitions. Examples considered include $\rho \to   \pi \pi$; $b_1 \to
\pi \omega$; hybrid meson decays and scalar meson  decays. I discuss 
what lattices can provide to help understand the composition of hadrons. 
\end{abstract}
 
 
 \maketitle

\section{Introduction}

  Relatively few hadronic states are stable to strong decays (i.e. via
QCD with degenerate $u$ and $d$ quarks).  Among the mesons, we
have~\cite{Eidelman:2004wy}:

\bigskip
\begin{tabular}{ll}
 { Stable} &{  $\pi\ K\ \eta\ D\ D_s\ B\ B_s\ B_c$}\\
 &  { $D_s^*\ B^*\ B_s^*\  D_s(0^+)\  B_s(0^+)$}   \\ 
 { $\Gamma < 1$ MeV} &{  
$\eta'\ D^*\ \psi(1S)\  \psi(2S)\ \chi_1\ \chi_2$} \\ & {$
\ \Upsilon(1S)\ \Upsilon(2S)\ \Upsilon(3S)$}\\ 
{  $\Gamma < 10$ MeV }&{  
$ \omega\ \phi\ \chi_0\ X(3872) $} \\
 {  $\Gamma > 10$ MeV }&{    $ \rho\ f_0\ a_0\ h_1\ b_1\ a_1\
f_2\ f_1\ a_2$}\\&{    etc., inc   $\eta_c$.}    
 \end{tabular}

\bigskip

The mass of an unstable state is usually defined as the energy
corresponding  to a 90$^0$ phase shift. This definition\footnote{Note
that defining the mass as the real part of the pole will cause a
downward shift  of masses for wider states, eg. 22 MeV
less~\cite{Michael:1966aa} for the $\Delta(1232)$ pole, and this
prescription  will fit the equal mass rule less well.}
 seems to accord with simple mass formulae:
  For example
 \begin{itemize}
 \item  $\rho(776)$ and  $\omega(783)$ are close in mass despite having 
widths of $150$ and  $8 $ MeV respectively.
 \item The baryon decuplet \\ ($\Delta(1232)$,  $\Sigma(1385)$,  $\Xi(1530)$,
 $\Omega(1672)$) is roughly \\ equally  spaced in mass despite having widths
of \\ (120, 37, 9, 0) MeV respectively.
 \end{itemize}

 So, on the one hand, unstable particles seem to  fit in well with
stable ones;  on the other hand, the presence of open decay channels
will have an influence in  lattice studies.

Some of the motivations to study hadronic decays on the lattice are:
\begin{itemize}
 \item What are hadrons made of? \\
 {\em  Is a meson made predominantly of  $\bar{q} q$, $\bar{q}\bar{q}qq$
or  meson-meson?}

 \item A state that can decay strongly (resonance) necessarily has 
a meson-meson component - is this important? \\
  { \em   Are unstable particles different from stable ones?}

 \item What is the nature of light-light scalar mesons? \\
 {\em   Where is the glueball?}

 \item  Are there hybrid mesons?

\end{itemize}

 Lattice QCD is a first-principles method of attack. But we use 
unphysical (too heavy) quark masses, we have Euclidean time.
 So what can we learn?

\section {Decays in Euclidean Time} 

  { NO GO.} At large spatial volume, the two-body continuum
{\em  masks}  any resonance state.
        The extraction of the spectral function from the correlator $C(t)$ is 
ill-posed unless a model is made~\cite{Michael:1989mf,Maiani:1990ca},
 since the low energy continuum dominates at large $t$.

\noindent    { GO}.   For finite spatial volume ($L^3$), the two-body
continuum is  {\em  discrete} and L\"uscher
showed~\cite{Luscher:1986a,Luscher:1986b,Luscher:1990ux} how to use the
small energy  shifts with $L$ of these two-body levels to extract the
elastic scattering phase shifts. The phase shifts then determine the
resonance mass  and width, see ref.~\cite{Luscher:1991cf} for a review. 
    { Thus a relatively broad resonance such as the $\rho$
appears as a distortion of the $\pi_n \pi_{-n}$  energy levels
where pion momentum $q=2 \pi n/L $.}

As a check of this approach to unstable particles on a lattice,  the
coupling of $\rho$ to $\pi \pi$ has been determined from first
principles~\cite{McNeile:2002fh}.
 The method is to arrange the $\rho$ and $\pi \pi$ state (with definite
relative momentum) to  be approximately degenerate in energy on a
lattice. 
 Then several independent methods allow to determine the transition
amplitude $x$ and, hence, the effective coupling constant $\bar{g}$ from
the lattice (where decay does not proceed)  and compare with experiment:

\bigskip
\begin{tabular}  {llll}
 \hline
  method  &     $m_{val}$  & $m_{sea}$ & $\bar{g}$ \\ 
 \hline
 { Lattice} $xt$ &         $s$   & $s$    &  $1.40^{+47}_{-23}$ \\ 
 { Lattice} $\rho$ shift &         $s$   & $s$    &  $1.56^{+21}_{-13}$ \\ 
 \hline
 $\phi \to K \bar{K}$ &         $s$   & $u,\ d$    &  $1.5$ \\ 
 $K^* \to K \pi$ &         $u,\ d/s$   & $u,\ d$    &  $1.44$ \\ 
 $\rho \to \pi \pi$ &         $u,\ d$   & $u,\ d$    &  $1.39$ \\ 
 \hline
\end{tabular}
\bigskip

  Note that the lattice has heavier sea quarks than experiment. Nevertheless, 
the level of agreement between first-principles lattice evaluation and 
experiment is very encouraging.

\subsection{Hadronic transitions from the lattice}

Here I summarise the steps that allow a fairly direct determination of
hadronic transition  amplitudes from the lattice.

{(i)} Consider a lattice study of the off-diagonal correlator: from a
$\rho$ meson to  $\pi \pi$. Diagrammatically:

\ \ \ \ \  {  $\rho$} $\to$ {  $\pi \pi$} \\
\ \ \ 0{  ----------}X{ ----------}0 \\
\ \ \ 0 \ \ \ \ \ \ \ \ \ \ \ t\ \ \ \ \ \ \ \ \ \ \ \ \ T

\includegraphics{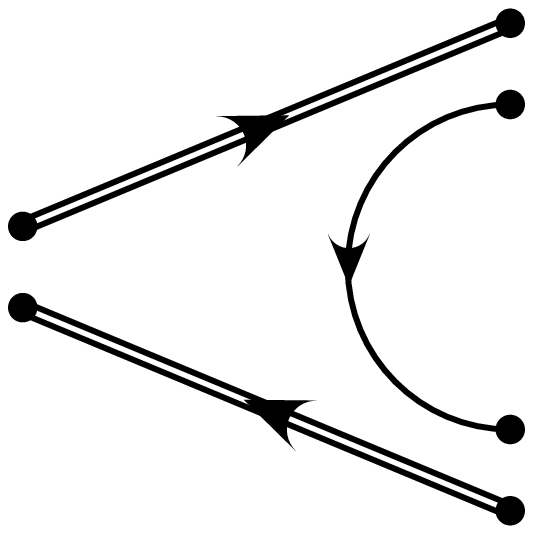}

{(ii)} Now to evaluate this contribution, since the
intermediate point marked X at time $t$ is not observed on a lattice, it
must be summed over:
   $$ \sum_{t=0}^T {  e^{-m(\rho)t}} x {  e^{-m(\pi \pi)(T-t)}}
 \to   xTe^{-mT} $$
 if ${  m(\rho)} \approx {  m(\pi \pi)}$ 

 {(iii)} So a plot of the (normalised) transition from the
lattice versus $T$ has  slope  of $x$, which can be
related~\cite{McNeile:2002fh} to the continuum coupling $g^2$. 

 \bigskip

Note this approach is like measuring the wave-function overlap - but
with no model  for the wave-functions.  In order to control excited state 
contributions, however, it is subject to the
restriction that initial and final states have  similar energies on a
lattice. A more rigorous approach is possible by using the method  
of determining the two-body energy versus lattice volume,
described above, but in practice sufficient precision is not available 
in general to study resonance decays, although results for 
scattering lengths have been obtained.

\section {Hybrid meson decay}

 One of the characteristic predictions of QCD is that there can be
mesons in which the gluonic degrees of freedom  are non-trivially
excited. The simplest example is  a hybrid meson with  spin-exotic
$J^{PC}=1^{-+}$ which is a $J^{PC}$ combination not available to  a
$\overline{q} q$ state. The spin-exotic quantum numbers then require 
a non-trivial gluonic contribution to the state. These hybrid mesons 
have been studied extensively on the lattice, here I discuss their decay.

\subsection{Light quarks}

\begin{figure}
\resizebox{0.50\textwidth}{!}{%
  \includegraphics{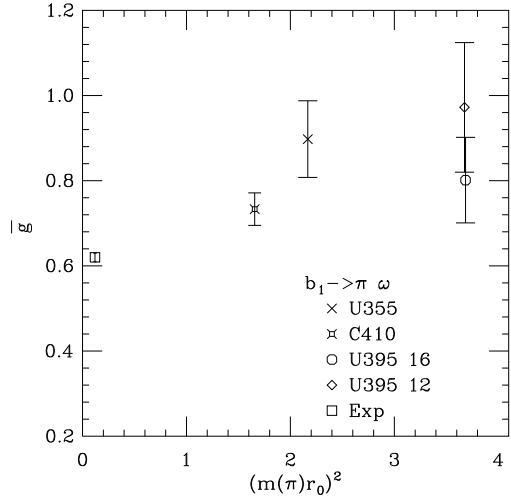}
}
 \caption{First test of  S-wave decays from the lattice for  $b_1 \to
\pi \omega$ from ref~\cite{McNeile:2006bz}. The effective coupling
$\bar{g}$ is plotted  versus the quark mass, as determined by the
pseudoscalar mass squared with the scale is set by $r_0 \approx 0.5$ fm.
 }
 \label{fig:lhyb.slope}
\end{figure}

 The S-wave decay of the  $J^{PC}=1^{-+}$ spin-exotic hybrid
meson($\hat{\rho}$) to $\pi b_1$ has been studied recently on the
lattice~\cite{McNeile:2006bz}. Since S-wave
 decays have not previously been studied in this way, a check was made
by  extracting the decay strength for the S-wave component of the
transition $b_1 \to \pi \omega$. As shown in fig~\ref{fig:lhyb.slope},
the lattice determination, though using heavier quarks than experiment,
fits in well. This gives confidence that the hybrid decay  prediction
will be reliable.
   Since the experimental effective coupling constant lies  lower than
the lattice results, this also shows that lattice results, with heaver
than  experimental quark masses, may overestimate the coupling somewhat.
This  may be interpreted, phenomenologically, as arising from
form-factor effects~\cite{Burns:2006wz}.

\begin{figure}
\resizebox{0.50\textwidth}{!}{%
  \includegraphics{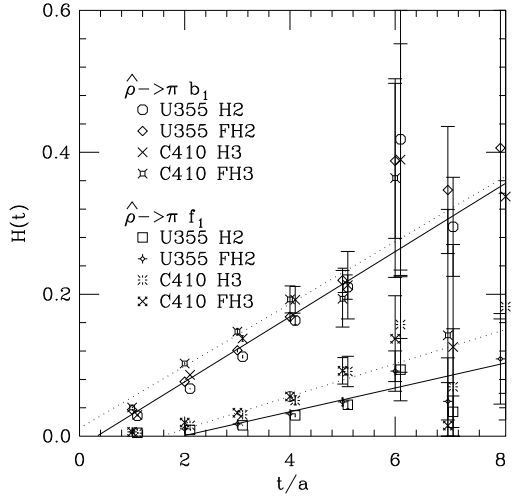}
}
 \caption{Strength of spin-exotic hybrid decay transition is given by the
slope from ref~\cite{McNeile:2006bz}.}
 \label{fig:lhyb.xt}
\end{figure}

    From lattices with $N_f=2$ flavours of sea-quark,  a recent
study~\cite{McNeile:2006bz} obtains a spin-exotic  hybrid meson state 
at 2.2(2) GeV. The S-wave decay transitions are then evaluated, as shown
in fig.~\ref{fig:lhyb.xt},   obtaining a partial width to $\pi
 b_1$ of  400(120) MeV and to  $\pi f_1$ of 90(60) MeV.
These results indicate that the decay width of this hybrid meson 
may be large and, hence, more difficult to extract experimentally.  

 For a recent preliminary lattice study of decay of a $J^{PC}=1^{-+}$
hybrid meson to $\pi a_1$  using the L\"uscher method which obtains a
width of around 60 MeV, see  ref.~\cite{Cook:2006tz}. This  decay
channel implies that the  hybrid meson considered has I=0, rather than
I=1 as above.

\subsection{Heavy quarks}

\begin{figure}[htb]
\resizebox{0.30\textwidth}{!}{%
  \includegraphics{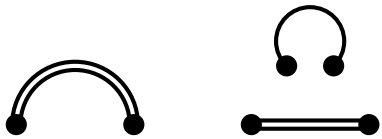}
}
 \caption{Initial and final states relevant for decay of heavy quark 
spin-exotic hybrid meson to $\chi_b f_0$.
 }
\label{fig:fd_hyb}

 \end{figure}

The cleanest environment in which to study such hybrid states on a
lattice is  in the limit of very heavy quarks which is relevant to
$\overline{b} b$. This can be approximated by using  static quarks and
the gluonic excitation arises as an excited string state  between these
static quarks with non-trivial gluonic angular momentum. Lattice studies
have long predicted the spectrum of such states. 

To guide experiment, however, it is important to know the expected decay
 mechanism and associated width. In the static quark limit, several 
symmetries can be used which imply~\cite{McNeile:2002az} that the
dominant  decay will be string de-excitation (rather than string
breaking) as illustrated in fig~\ref{fig:fd_hyb}. Lattice
study~\cite{McNeile:2002az} shows that the  dominant decay of the hybrid
meson $H_b$  is string de-excitation  to  $\chi_b f_0$.   The  width is
predicted to be around 80 MeV.  

 This  estimate from first principles of the decay width is of 
significance in guiding experimental searches for such hybrid states.


\section {Scalar Mesons}

\subsection{Light quarks}

 Since $u\overline{u}+d\overline{d}$, $s\overline{s}$,  glueball, and
meson-meson components are all possible for flavour-singlet scalar
mesons, this is  a difficult area to study both on a lattice, and in
interpreting experimental data.  For scalar mesons the lowest  mass
decay channels are $\pi \pi$ (flavour singlet: $f_0$)  or $\eta \pi$
(flavour non-singlet: $a_0$) and these decay channels are open in many
dynamical lattice studies. The history of lattice attempts to study the
complex mixing between these  different contributions is :

 \begin{itemize}
   \item  $0^{++}$ glueball decay $\to \pi \pi$: quenched
study~\cite{Sexton:1995kd,Sexton:1996ed}.
   \item glueball mixing with $q \overline{q}$ meson. 
 This   hadronic transition  has been studied using
quenched~\cite{Lee:1999kv} and dynamical lattices~\cite{McNeile:2000xx} 
  \end{itemize}

      A full lattice  study is needed which includes glueball,
$\overline{q} q$ and $\pi \pi$  channels but the disconnected diagram
for $f_0 \to \pi \pi$  is very noisy in practice - as shown in 
ref.~\cite{cmcmjp}.

\begin{figure}[htb]
\resizebox{0.50\textwidth}{!}{%
  \includegraphics{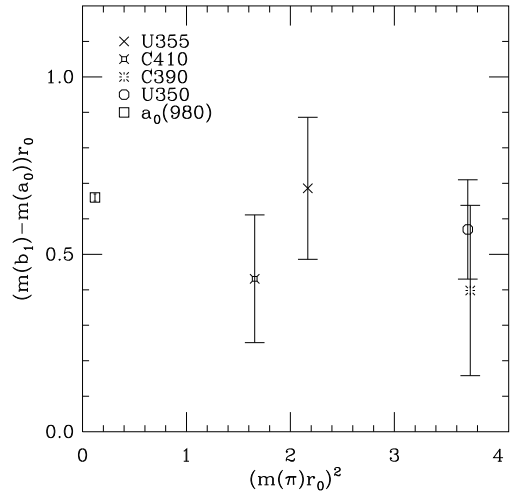}
}
 \caption{Mass difference of $b_1$ and $a_0$ mesons from
ref~\cite{McNeile:2006nv} plotted  versus the quark mass, as determined
by the pseudoscalar mass squared with the scale is set by $r_0 \approx
0.5$ fm. The open box is the experimental point if the $a_0$ meson is at 
980 MeV.
 }
\label{fig:mba}

 \end{figure}

 To reduce the contribution from disconnected diagrams, one can study
flavour non-singlet scalar mesons. 
 The simplest case is $a_0$ which has a decay $ \eta \pi$ and this has
been explored in quenched studies  which have an anomalous behaviour:
since the $\eta$ itself is unphysical (appearing as a double pole
degenerate in mass with the pion). Rather than try to correct for this
anomaly which gives  a wrong sign to the $a_0$ correlator at larger $t$,
it is preferable to use  a ghost-free theory. 
 With two flavours of sea quark ($N_f=2$), this problem is avoided.  A
recent study~\cite{McNeile:2006nv} of the $a_0$ meson concentrates on
the mass difference between it and  the $b_1$ meson, as shown in 
fig.~\ref{fig:mba}. This study concludes that the $\bar{q}q$ non-singlet
 scalar meson lies around 1 GeV - which is considerably lighter than
some previous lattice studies (see ref~\cite{McNeile:2006nv} for a
summary).

\begin{figure}[htb]
\resizebox{0.50\textwidth}{!}{%
  \includegraphics{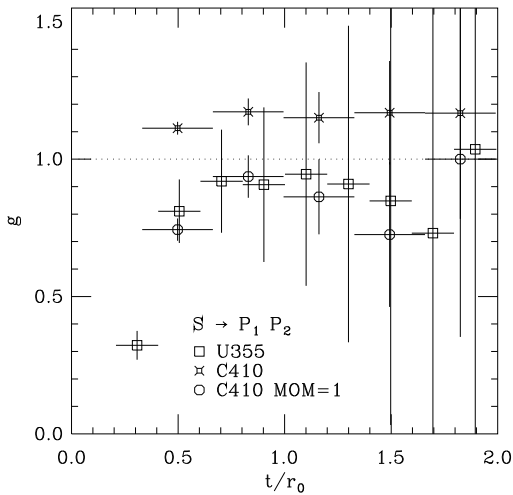}
}
 \caption{Effective coupling from connected contribution to  decay
transition of scalar meson $\to$ 2 pseudoscalars from
ref.~\cite{McNeile:2006nv}. The $t$ region around 0.5 to 1.5 is expected 
to be relevant (in units  with $r_0 \approx 0.5$ fm).
 }
\label{fig:a0f}

 \end{figure}

 As well as determining the mass values, this study evaluates the 
 $a_0 \to \eta \pi$ and $a_0 \to K K $  transitions. Results from the
connected contribution to these  transitions are shown in
fig.\ref{fig:a0f}. The resulting coupling constant is determined to be
of similar value to that obtained by some phenomenological 
studies of decays of the $a_0(980)$ meson.
 This again points to the possibility that the $a_0(980)$ may be
substantially  a $\bar{q}q$ state.

 A full study of flavour singlet scalar mesons will need to take into account 
the $\bar{q}q$, gluonic and meson-meson channel and their mixing. This has 
not yet been achieved, for  a summary of the current state of lattice studies 
see ref.~\cite{cmcmjp}.

 \subsection{Heavy-light quarks}

 One of the most promising ways to study scalar mesons on  lattice is
through  heavy-light mesons. The scalar meson with $\overline{c}s$ quantum
number is known  experimentally~\cite{Eidelman:2004wy} to be very narrow
(it decays only via the isospin-violating  channel $D_s \pi$ or
electromagnetically), while the scalar meson with  $\overline{b}s$
quantum number is predicted to be similarly narrow from a lattice study 
of its energy~\cite{Green:2003zz}. 

\begin{figure}[htb]

\resizebox{0.30\textwidth}{!}{%
  \includegraphics{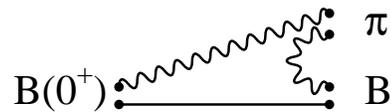}
}
 \caption{ Diagram for $B(0^+) \to B  \pi$ transition
 }
 \label{fig:BsBpi}
 \end{figure}

\begin{figure}[htb]

\resizebox{0.50\textwidth}{!}{%
  \includegraphics{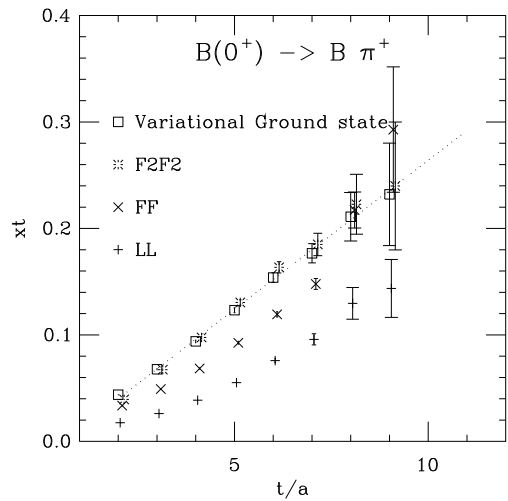}
}
 \caption{$B(0^+) \to B  \pi$  transition strength (given as slope) on a
lattice from ref.~\cite{McNeile:2004rf}.
 }
 \label{fig:xt_bs}
 \end{figure}

 The heavy-light scalar meson, which in the limit of a static heavy
quark, is  expected to be stable~\cite{Green:2003zz} for $\overline{b} s$
content and to decay to $B \pi$ for  $\overline{b}n$ content (where $n=u,\
d$, considered as degenerate).  
 Evaluating the diagram shown in fig.~\ref{fig:BsBpi}, a lattice
estimate, shown in fig.~\ref{fig:xt_bs}, of the decay rate  of  $B(0^+)
\to B(0^-) \pi$   gives a width predicted~\cite{McNeile:2004rf}  as 
162(30) MeV. This state has not been  observed experimentally yet, but
the  experimental results for the corresponding $\overline{c}n$ state,
$D(0^+)$,  are that the width is  $270\pm50$ MeV. Although significant 
$1/m_Q$ effects are expected in the HQET in extrapolating to charm
quarks, this is indeed a similar magnitude to that predicted for $B$
mesons. 
 It will be interesting so see how the lattice prediction of the  mass
and width of the $B(0^+)$ fares when experimental results are available.

\section {Do decays matter?}

  The previous discussion of decays on a lattice emphasises  that $q
\overline{q}$ states do mix with two-body states  with the same quantum
numbers. In the real world, the two-body states are a continuum  and
nearby states have a predominant influence.  Since there is a
suppression in the amplitude near threshold from  the factor $q^L$ for
an $L$-wave transition,  for S-wave transitions ($L=0$)  the threshold
will turn on most abruptly and hence will have stronger mixing.

For {  bound} states there is an influence of nearby many-body states
(eg. $N \pi$ on $N$ or $\pi \pi \pi$ on $\pi$) which mix to reduce the
mass. The nearest such  thresholds will be those with pionic channels
since pions are the lightest mesons.  This is the province of low energy
effective theories, especially Chiral Perturbation Theory which is
discussed in other  talks. This then provides a reliable guide in
extrapolating lattice results to the  physical light quark masses.

For {  unstable} states (resonances) the influence of the two-body
continuum  is less clear since the two-body states are both lighter and
heavier.  In the continuum at large volume, effective field theories 
can again be used to explore this. On a lattice, however, the signal for
 a particle becomes obscured as the quark mass is reduced so that it
becomes unstable. Techniques, such as those discussed above, are needed to 
extract the elastic scattering phase shift and hence the mass and width.

For quenched QCD, however, where these two-body states  are not coupled
(or have the wrong sign as in $a_0 \to \eta \pi$), then the  unstable
states will be distorted compared to full QCD. For instance, in existing
quenched QCD studies,  the  $\rho$ will be {\em too heavy} since it is
not repelled by the  heavier $\pi \pi$ states. Indeed an example of this
effect was seen above in the study of $\rho$ decay including dynamical
sea quarks, where the  $\rho$ mass decreased~\cite{McNeile:2002fh} when
it could couple to  $\pi \pi$ compared to  when it could not.

\section {Molecular states?}

 Can lattice QCD provide evidence about possible molecular states: 
hadrons made predominantly of two hadrons?

 The prototype is the deuteron: $n\  p$ bound in a relative S-wave (with
some D-wave admixture) by $\pi$ exchange.

 There are many states close to two-body thresholds. Since S-wave
thresholds are the  most abrupt, it is usually in this case that  the
influence of the  threshold on the state has been discussed. 
Some of these   cases are:

\bigskip
 \begin{tabular}{rcl}
$f_0(980)\ a_0(980)$ &$\leftrightarrow$& $K \overline{K}$ \\
       $ D_s(0^+)$ &$\leftrightarrow$& $D(0^-) K$\\
       $ B_s(0^+)$ &$\leftrightarrow$& $B(0^-) K$ \\
       $ X(3872)$ &$\leftrightarrow$& $D^* \overline{D}$\\ 
       $ \Lambda(1405)$ &$\leftrightarrow$& $\overline{K} N$ \\
       $ N(1535)$ &$\leftrightarrow$& $\eta N$\\
 \end{tabular} 
\bigskip

Some of these states ($ D_s(0^+)$, $B_s(0^+)$) are stable (in QCD in the
isospin  conserving limit) whereas the rest have other channels open.  
There is a large literature, stretching over 40 years, discussing the 
consequences of the nearby threshold on these states. One definite 
implication is that  isospin breaking is enhanced by mass splittings in
thresholds (eg. $\overline{K^0}  K^0$ compared to $K^+ K^-$ is 8 MeV higher 
and this induces isospin mixing between the states at 980 MeV). 
This level of detail is not accessible in lattice studies at present, but 
lattice QCD should be able to address the issue of the influence of 
thresholds on these states.


 The observation of a state near a  2-body threshold implies that there 
is an attractive interaction between the two bodies.  But this is a
topic like that of whether the chicken or egg was created first:  an
attractive interaction implies and is implied by a nearby state. What
can lattice QCD offer here? We are in the position of being able to 
vary the quark masses and this is a very useful tool. A two-body
threshold will move in general in a different way with changing quark
mass than a $\overline{q}q$ state. We can also move the strange and
non-strange masses  separately and this can be helpful too.

Another line of investigation is that lattice studies can explore the 
wavefunction of a state - either the Bethe-Salpeter wavefunction or the 
charge or matter spatial distribution.  One can also explore the
coupling of a state to a 2-body channel, as was discussed  above. 

The prototype of a molecular state is the deuteron: it has a tiny
binding energy (2.2 MeV)  and a very extended spatial wave function.
Pion exchange between neutron and proton  gives a mechanism for this
long-range attraction. In general it is difficult to  reproduce such
small binding energies in lattice studies. 

Another case where a long-range pion exchange can give binding is in the
$BB$  system. Here lattice results indicate~\cite{Michael:1999nq} the
possibility  of molecular bound   states in some quantum number channels
which have  an attractive interaction from pion exchange,  but also the 
possibility of bound multi-quark states which are not described as
hadron-hadron  but where the two heavy quarks form a colour triplet and
the light quarks  are arranged as in a heavy-light-light baryon. This
$BB$ example illustrates the rich structure available to multi-quark 
systems.

One case where lattice studies have been able to shed considerable light
is  for the $\overline{b}s$ and $\overline{c}s$  scalar mesons. Since,
in the isospin conserving limit, these  states cannot couple to $B_s \pi$,
$D_s \pi$, they have a lightest open decay channel  $BK$, $DK$. Lattice
studies~\cite{Green:2003zz,Dougall:2003hv} indicate that, in both cases,
the scalar meson lies below the open threshold, so the states should be
stable. 

\begin{figure}[htb]

\rotatebox{270} {%
\resizebox{0.35\textwidth}{!}{%
  \includegraphics{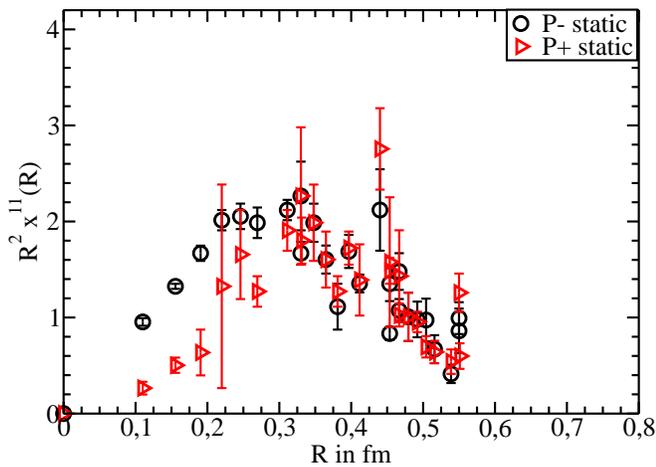}
}}
 \caption{Charge distributions of $B_s(0^+)$ and $B_s(2^+)$ mesons from
the lattice. The heavy quark  is static and the light quark is at
distance $R$ from it. The  $J^P=0^+$ meson is $P_-$ while the $J^P=2^+$
meson is $P_+$. 
 }
 \label{fig:jk2}
 \end{figure}

Because simple quark model expectations were that these  scalar mesons
were unstable, theorists have suggested that a $BK$, $DK$  molecular
composition was responsible. This can be explored on a lattice by
measuring  the spatial distribution of the heavy-light meson.  A
study~\cite{Green:2005st}  of the charge  distribution of the light
quark in a $B_s(0^+)$  meson is illustrated  in  fig.\ref{fig:jk2}.
This shows that the light quark spatial distribution is similar to that
of other  $\overline{Q}q$ states (eg. $J^P=2^+$) for which no  molecular
interpretation is proposed.   This reinforces the conclusion that the
$B_s(0^+)$ is predominantly  a $\bar{b}s$ state. The decay transition
from  $B_s(0^+)$ to $BK$ has also been determined (see above) and it has
an effective coupling constant which is consistent  with that found for other 
(non-molecular) decays. Overall, lattice evidence does not support 
the hypothesis that $B_s(0^+)$ is a molecular state.

\section {Conclusions}

 Lattice can address hadronic structure:  
 \begin{itemize}
 \item form factors (eg. charge wavefunctions) can be evaluated
 \item decay transitions (and mixing transitions) can be evaluated
 \item structure function moments can be evaluated (steady but slow
progress here)
 \item hadronic matrix elements are needed to interpret experiment 
(eg $f_B$ relates $B$ meson to $b$ quark, etc..) in searches 
for signs of physics beyond the standard model.
 \end{itemize}

 Hadronic physics involves {\em unstable} states and
 lattice techniques are being developed to  study these as we have
summarised.
 These techniques have been tested against experiment for $\rho \to \pi
\pi$ and $b_1 \to \pi \omega$. 
 In particular we presented evidence that the  spin exotic hybrid meson
(made of light quarks) is at a mass around 2 GeV and has a wide width.
   We discussed scalar mesons, and presented  evidence that the 
$a_0(980)$ meson  is basically a $q \bar{q}$ state and  that the
$B_s(0^+)$ meson  is also predominantly a $b \bar{q}$ state.

 There is a lot to be learnt from the lattice beyond mass spectra.

%

\end{document}